\documentclass[apj,twocolumn]{oja}
\usepackage{amsmath, amssymb, amsfonts}
\usepackage{graphicx} 
\usepackage{xcolor}
\usepackage{hyperref}

\usepackage{appendix}
\usepackage{comment}
\usepackage{natbib}
\hypersetup{colorlinks=true,
    citecolor=blue,
    linkcolor=blue,
    }

\begin{document}

\title{Inverting Fisher biases for fast systematics exploration}
\author{Biancamaria Sersante}
\author{Christos Georgiou}
\author{Nora Elisa Chisari}
\email{$\dagger$ sersante@strw.leidenuniv.nl}
\affiliation{Leiden Observatory, Leiden University, PO Box 9513, 2300 RA Leiden, the Netherlands.}
\affiliation{Institut de Física d’Altes Energies (IFAE), The Barcelona Institute of Science and Technology, Campus UAB, 08193 Bellaterra (Barcelona), Spain}
\affiliation{Institute for Theoretical Physics, Utrecht University, Princetonplein 5, 3584 CC, Utrecht, the Netherlands.}

\begin{abstract}

Upcoming cosmological surveys will achieve increasingly precise constraints in cosmological parameter estimation. To guarantee the robustness of cosmological analyses, it is essential to account for and model systematic effects that can bias cosmological constraints, shifting the best fit parameters away from their fiducial values. It is possible to approximately infer the biases that un-modelled systematic effects might introduce in cosmological parameter estimation by means of the Fisher matrix formalism. In this paper, we introduce a new application of this formalism, where by inverting the process, we investigate whether a specific missing or mis-modelled systematic effect can explain away a given tension between two different probes or experiments. We showcase the proposed methodology by examining two representative systematics: galaxy intrinsic alignments and baryonic feedback. As the method is agnostic to the systematic effect and can be applied to a wider range of scenarios, we discuss more possible future applications.
While the proposed approach is accurate in the limit of small offsets in the cosmological parameters, where the likelihood can be considered linear in both the cosmological parameters and the systematics, in practice, the region of validity depends on the systematic effect. 
In general, even beyond this region, the approach still provides a useful test that helps indicate the magnitude and direction of potential biases from systematic effects in data.  
\end{abstract}

\maketitle

\section{Introduction}

Between the late 20th and the early 21st century,
the discoveries of the cosmic microwave background (CMB), the large-scale structure, and  distance measurements from supernovae \citep[see e.g.][]{PenziasWilson1965,DavisPeebles1983,Riess,Perlmutter} consistently indicated that the Universe can be described by a standard cosmological model: the $\Lambda$-Cold Dark Matter ($\Lambda$CDM) model.
However, the increased constraining capability of recent experiments has revealed some tentative remaining tensions between cosmological data sets \citep[see e.g.][]{Efstathiou2025,DESIVII2025,DESBAO2025} that, in the absence of unaccounted for systematic errors, might imply the need to extend the $\Lambda$CDM model. 

In light of this, the goal of the next generation of observational (Stage-IV) campaigns \citep[see e.g.,][]{SRD, Euclid2025} is to achieve sub percent-level precision in the measurement of the matter distribution in the Universe and the growth of the density perturbations between $0<z<3$. Such effort can determine whether tentative tensions seen in the current generation of Stage-III experiments remain and could be adjudicated to deviations from $\Lambda$CDM.
The quality and precision of these experiments can allow for a dramatic reduction of statistical errors but, to correctly assess the final uncertainties, it is fundamental to also maintain biases in cosmological parameter constraints from systematic effects below the statistical errors. 

In fact, the key to the successful design of upcoming experiments lies in our ability to predict the sensitivity of the measurements to modelling uncertainties and observational systematics, and how these propagate into the cosmological parameters \citep{Mandelbaum2018, Euclid2025}. More specifically, the goal is to optimize decisions to be taken (e.g., about modelling choices) during analysis in order to extract as much cosmological information as possible in the quickest and most robust way.

The Fisher information matrix has become a widely used tool in cosmology for predicting how well cosmological parameters are expected to be determined in a given experiment if the likelihood is Gaussian \citep[see e.g.,][]{Tegmark1,Tegmark2}. 

The Fisher formalism has also been extended to predict offsets in the best-fit parameter space when known, unaccounted systematics are present \citep[see e.g.,][]{Taylor,Amara,Kitching, Duncan,  Takada,Bernal}.
For instance,  \cite{Takada} considered the impact of photometric calibration errors and power-law shape systematics on measurements of cosmic shear when constraining the matter density in the universe $\Omega_{\rm m}$, the amplitude of mass fluctuations $\sigma_8$ and the dark energy equation of state parameters $w_0, w_{\rm a}$  (namely, $w(z) = w_0 + w_{\rm a} z/(1 + z)$, according to \citealt{Linder} and \citealt{Chevallier}) using the Fisher matrix formalism. A similar approach was taken by \cite{Huterer} to study the impact of theoretical uncertainties in modelling the matter power spectrum with $N$-body simulations. \cite{Amara} extended the Fisher matrix analysis to include a calculation of bias to study the impact of residual systematics on tomographic cosmic shear surveys.
\cite{Bernal} generalized previously proposed expressions to estimate a priori systematic errors and illustrated the relevance of the method in the case of multi-tracer analyses of galaxy clustering.

In this paper, we present a new application of the Fisher bias formalism, employing it to investigate tensions between constraints on cosmological parameters resulting from different experiments. Typically, when tensions arise, many expensive likelihood sampling algorithms need to be re-run to study whether un-modelled systematic effects could explain the observed data tensions. Here we show how this can be avoided using the Fisher formalism to derive the expected parameter values of known systematic effect that could explain the bias in cosmological parameters obtained from a given experiment.
In fact, Fisher biases indicate both the magnitude and the direction in which an un-modelled systematic effect shifts the best-fit parameters. This relation can be inverted allowing us to determine whether a missing known systematic effect can explain an observed tension between different experiments by estimating the value of the systematic parameter which would generate the observed discrepancy in a computationally inexpensive way.

We showcase the new methodology by providing different examples of measuring and modelling of galaxy shapes. Weak gravitational lensing changes the observed shapes of galaxies by distorting the path of photons along the line of sight \citep[see e.g.,][]{BartelmannS2001, Hoekstra2008}. These distortions can be extracted from imaging data and modelled to infer cosmological parameters in the $\Lambda$CDM model. However, several systematic effects need to be carefully considered when inferring cosmological information from weak lensing, and their mitigation is crucial to avoid biased parameter constraints. The most important astrophysical systematic effects are the intrinsic alignments of galaxies \citep[][]{Troxel2015,Kirk2015,2015SSRv..193...67K,Joachimi15,Chisari2025} and the effect of baryons on the matter power spectrum \citep{vanDaalen11,ChisariBaryons,2025PhRvL.134h0001C}. We demonstrate the Fisher bias application by performing a mock analysis of weak lensing Stage-IV data while neglecting intrinsic alignments or baryonic feedback in the modelling, to emulate a bias in cosmological parameters that would generate observed tensions in the literature.

This paper is organised as follows. In Section \ref{section2}, we summarize the methodology to express the bias in terms of Fisher matrices and derive an equation that relates such bias to systematic parameters. In Section \ref{newappl} we derive the expression of our inverted relation.
In section \ref{implementation}
we first introduce our forecasting framework (including the redshift distribution used for the galaxy-sample and predictions for the angular power spectra).
In Section \ref{exploration}, we showcase possible applications of the inverted bias relation, based on modelling intrinsic alignments and baryonic feedback.
Finally, in Section \ref{discussion} we summarize and discuss our findings. 

\section{Parameter bias estimation}\label{section2}
\subsection{Fisher formalism}

The procedure to obtain estimates of the values that free parameters $\boldsymbol{\theta}=(\theta_1,...,\theta_m)$  should have in a certain model $M$ to fit the data $\mathbf{d}=(d_1,...,d_n)$ with covariance $C$, is referred to as \emph{parameter estimation} \citep[see e.g.,][]{Tegmark2,Dodelson, Trotta2008}.

In general, the quantity we are interested in is the probability distribution function of a certain set of cosmological parameters in a model given the data. According to Bayes Theorem, this can be expressed as
\begin{equation}
P\left(M \mid\mathbf{d}, C\right)=\frac{P\left(\mathbf{d}, C\mid M\right) \, P\left(M\right)}{P\left(\mathbf{d}, C\right)},
\end{equation}
where:
\begin{enumerate}
    \item $P\left(\mathbf{d}, C\mid M\right) \equiv \mathcal{L}\left(\mathbf{d}, C\mid M\right)$ is the likelihood of the data given the model;
    \item $P\left(M\right)$ is called the prior and encodes knowledge gained from theoretical constraints and previous experiments. 
    We point out that final results should not be very sensitive to the prior, otherwise the data would not be constraining enough.
    \item $P\left(\mathbf{d}, C\right)$ is called the evidence and it does not influence the posterior distribution directly as it acts as a simple normalisation factor. The evidence is very useful when comparing different models using the same data, because the ratio between the evidence of different models provides information on how strongly the data favor one model over another.
\end{enumerate}

Maximizing the posterior probability, when results are insensitive to the prior, amounts to maximizing the likelihood. As we assume the likelihood functions we are interested in to be Gaussian, it is most convenient to maximize their logarithm; thus, we have to find the parameters $\hat{\theta}_i$ (\emph{best-fit parameters}) such that
\begin{equation}\label{estimator_L}
    \dfrac{\partial\ln\mathcal{L}}{\partial\theta_i}(\mathbf{d}, C\mid M)\Big|_{\hat{\theta}_i}=0.
\end{equation}
Maximizing the logarithm of the likelihood is equivalent to minimizing $\chi^{2}$ (where $\chi^{2}\equiv -2\ln\mathcal{L}$).

In order to predict confidence intervals for the constrained parameters $\hat{\theta}_i$, we introduce the Fisher formalism, which allow us to anticipate how well we expect cosmological parameters to be determined in a given experiment. 
We assume a fiducial set of parameters, which are close to the true value of these parameters, and expand $\chi^2$ around the best-fit values as
\begin{equation}
\chi^{2}(\theta_i)=\chi^{2}(\hat{\theta}_i)+\frac{1}{2}\frac{\partial^{2}  \chi^{2}}{\partial \theta_{i} \partial \theta_{j}}(\theta_i-\hat{\theta}_i)(\theta_j-\hat{\theta}_j).
\end{equation}
To obtain confidence intervals for the model parameters, we can ask how quickly the likelihood changes as a certain parameter $\theta_{i}$ moves away from the corresponding best-fit value. If $\chi^{2}$ increases rapidly, the error on the parameter under consideration would be small, and vice versa.
With this heuristic explanation we understand that the elements of the Fisher matrix, defined as
\begin{equation}\label{Fishdef}
    F_{i j}=\langle \mathcal{F}_{ij}\rangle \equiv-\frac{1}{2}\left.\left\langle\frac{\partial^{2}  \chi^{2}}{\partial \theta_{i} \partial \theta_{j}}\right\rangle\right|_{\left\{\theta_{k}\right\}=\left\{\hat{\theta}_{k}\right\}},
\end{equation}
quantify how much information  on the model parameters can be extracted from the data of a given experiment. 
The average in the above equation is the average of all possible realisations of the model and it is derived under the assumptions of a Gaussian likelihood and a covariance independent of the cosmological parameters. 

In the next section, after briefly summarising how the Fisher formalism can be employed to estimate parameter bias, we analyse an extension of this formalism that can save computational resources in the exploration of whether a particular systematic effect might be responsible for a tension between data sets.
\section{Inverting the Fisher bias relation}\label{newappl}

We start by considering a data set $D$ that we try to fit with a model $M_1$ (with one free parameter $\theta$). Moreover, we consider some external data $E$, independent of $D$, which we do not use to obtain constraints on parameter $\theta$ (e.g., $E=$CMB vs. $D=$weak lensing). 
We also define an expanded version of the above model, denoted by $M_2$, which includes an additional systematic parameter $\psi$. In particular, we choose $M_1$ to be a subset of $M_2$ with $\psi$ fixed at a certain specific value, i.e., $M_1=M_2(\theta,\psi_{\text{fixed}})$.

Our goal is to check whether we can explain an observed shift in the cosmological parameter $\theta$ preferred by $D$, compared to $E$, by considering $M_2$ instead of $M_1$. 
The likelihood in $M_1$ is maximized at $(\theta_{\text{bias}},\psi_{\text{fixed}})$ (by construction in the case of $\psi$, as it has been fixed to a certain value different from its fiducial value, $\psi_\mathrm{fid}$).
\begin{figure}
    \centering
    \includegraphics[width=\columnwidth]{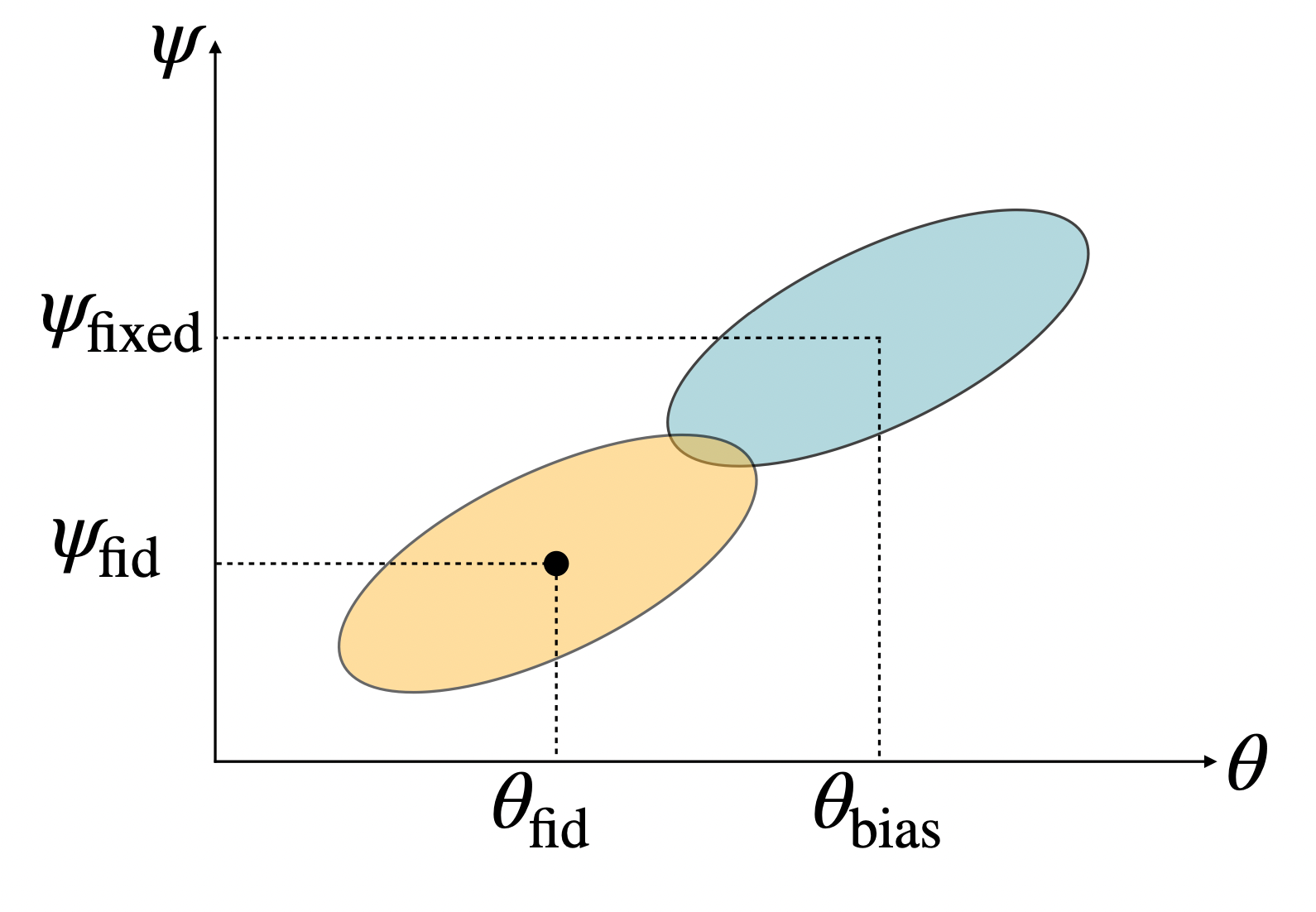}
    \caption{Cartoon representation of the confidence ellipses for the two models; in orange, the contour for model $M_2$, assumed to maximize the likelihood at the true values of the parameters under consideration ($\theta_{\rm fid},\psi_{\rm fid}$). In blue, the contour for model $M_1$, which, at a fixed value of $\psi$, maximizes the likelihood in $\theta_{\text{bias}}$. The shifts between the best-fit parameters in the two models, on the $x$- and $y$- axes, represent the biases $\delta \theta$ and $\delta \psi$, respectively.  }
    \label{Setup}
\end{figure}
We illustrate this set-up in Figure \ref{Setup} by showcasing the expected best-fit parameter values and corresponding contours for the two models considered. 
As model $M_1$ favours $\theta_{\text{bias}}$, for it to have a maximum at $\theta_{\text{bias}}$, it must be true that the likelihood peaks at this particular point in parameter space, that is

\begin{equation}\label{M1}
\left.\partial_\theta \ln \mathcal{L}_{M_1}(\theta)\right|_{(\theta_{\text{bias}},\psi_{\text{fixed}})}=0,
\end{equation}
where we denote the partial derivative with respect to a parameter $\theta$ as $\partial_\theta$ for brevity in the following calculations. 
Moreover, when fitting $M_2$ to the data $D$, the likelihood peaks at the fiducial values of the parameters:

\begin{align}\label{partial}
& \left.\partial_\theta \ln \mathcal{L}_{M_2}(\theta, \psi)\right|_{\left(\theta_{\text {fid }}, \psi_{\text {fid }}\right)}=0,\\
& \left.\partial_\psi \ln \mathcal{L}_{M_2}(\theta, \psi)\right|_{\left(\theta_{\text {fid }}, \psi_{\text {fid}}\right)}=0.
\end{align}
Given the following quantities,
$\Delta \theta \equiv \theta_{\text{bias}} - \theta_{\text{fid}}$ and $\Delta\psi \equiv \psi_{\text{fixed}} - \psi_{\text{fid}}$,
the questions we aim to answer are the following: 
\begin{enumerate}
    \item by fixing the value of $\psi$ in model $M_1$, can we obtain a shift $\Delta\theta$ compatible with the observed one?
    \item For the shift in $\theta$ we observe, is the value of $\Delta\psi$ reasonable compared to our expectations (e.g., from prior experiments)?
\end{enumerate}

To answer the above questions, we take the following approach.
By assuming that such shifts are small, we can Taylor expand $\partial_\theta \ln \mathcal{L}_{M_2}(\theta, \psi)$ around $\left(\theta_{\text {fid }}, \psi_{\text {fid }}\right)$ as 
\begin{align}
\left.\partial_\theta \ln \mathcal{L}_{M_2}(\theta, \psi)\right|_{(\theta, \psi)} &\simeq 0+\left.(\theta-\theta_{\rm fid}) \frac{\partial^2 \ln \mathcal{L}_{M_2}}{\partial \theta^2}\right|_{\left(\theta_{\text {fid }}, \psi_{\text {fid }}\right)}+\nonumber\\
&+\left. (\psi-\psi_{\rm fid}) \frac{\partial^2 \ln \mathcal{L}_{M_2}}{\partial \theta \partial \psi}\right|_{\left(\theta_{\text {fid }}, \psi_{\text {fid }}\right)}.
\end{align}
Here, the zero-order term in the expansion vanishes according to Eq. (\ref{partial}).
We recall that we are choosing  $M_1$ to be a subset of $M_2$ with $\psi$ fixed at a certain specific value.  
This translates into $\ln \mathcal{L}_{M_1}(\theta)=\ln \mathcal{L}_{M_2}\left(\theta, \psi_\mathrm{fixed}\right)$.
Therefore, if Eq. (\ref{partial}) is satisfied,  Eq. (\ref{M1}) must be as well.
As a consequence, the LHS of the above equation vanishes, resulting in
\begin{equation}\label{biasElisa}
\left.\Delta\theta \frac{\partial^2 \ln \mathcal{L}_{M_2}}{\partial \theta^2}\right|_{\left(\theta_{\text {fid }}, \psi_{\text {fid }}\right)}=-\left.\Delta\psi \frac{\partial^2 \ln \mathcal{L}_{M_2}}{\partial \theta \partial \psi}\right|_{\left(\theta_{\text {fid }}, \psi_{\text {fid }}\right)}.
\end{equation}
By recalling the definition of the Fisher matrix according to Eq. (\ref{Fishdef}), we can rewrite Eq. (\ref{biasElisa}) as follows
\begin{equation}\label{1parameterbias}
    \Delta\theta=-(F^{\theta \theta})^{-1}F^{\theta\psi}\Delta\psi.
\end{equation}
The above expression allows to compute the bias introduced in a cosmological parameter by the presence of residual (unmodelled) systematics using the Fisher matrix \citep[for further details on possible applications see,][]{Amara,Taylor,Bernal}. 
We note that, mathematically, the derivation is rigorous only for small deviations, where both 
$\Delta\theta$ and $\Delta\psi$ are \emph{small}. In practice, the formalism can provide useful guidance even for larger observed shifts for (a) smooth and well-behaved likelihoods, (b) systematics that are linear in the data vector (but not necessarily in the cosmological parameters), (c) any kind of systematics (linear or not) but provided that the offset is small.
In the next sections, we generalize Eq. (\ref{1parameterbias}) and from its general form, we develop our own application.


\subsection{Generalisation to multiple parameters}

In the previous paragraph we derived an expression for the bias induced by a single systematic parameter on a given cosmological parameter in the underlying model. In this section, we generalize the previous result to include $m$ cosmological parameters, $\boldsymbol{\theta}=(\theta_1,...,\theta_m)$, and $n$ systematics, $\boldsymbol{\psi}=(\psi_1,...,\psi_n)$. 

We start by considering a likelihood function $\ln \mathcal{L}(\boldsymbol{\theta}, \boldsymbol{\psi})$.
Analogously to the previous case, we have that, when fitting $M_2$ to the data, the likelihood peaks at the true parameter values, meaning that we can generalize Eq. (\ref{partial}) as
\begin{equation}\label{Vanish}
\left.\partial_{\theta_i} \ln \mathcal{L}_{M_2}(\boldsymbol{\theta}, \boldsymbol{\psi})\right|_{\left(\boldsymbol{\theta}_{\text {fid}}, \boldsymbol{\psi}_{\text {fid }}\right)}=0
\end{equation}
where $\partial_{\theta_i}$ is the partial derivative in the $\boldsymbol{\theta}$-parameter space with respect to $\theta_i$. Now, by Taylor expanding and taking the average over all possible realisations of the model, we obtain

\begin{align}
\begin{split}\label{generalder}
\left\langle\partial_{\theta_i} \ln \mathcal{L}(\boldsymbol{\theta}, \boldsymbol{\psi})\right\rangle &=\left.\left\langle\partial_{\theta_i} \ln \mathcal{L}(\boldsymbol{\theta}, \boldsymbol{\psi})\right\rangle\right|_{\left(\boldsymbol{\theta}_{\text {fid }}, \boldsymbol{\psi}_{\text {fid }}\right)} +\\
& +\left. (\theta_j-\theta_{\text {fid},j})\left\langle\partial_{\theta_i} \partial_{\theta_j} \ln\mathcal{L}(\boldsymbol{\theta}, \boldsymbol{\psi})\right\rangle\right|_{\left(\boldsymbol{\theta}_{\text {fid }}, \boldsymbol{\psi}_{\text {fid }}\right)}+ \\
& +\left.(\psi_j-\psi_{\text {fid }, j})\left\langle\partial_{\theta_i} \partial_{\psi_j} \ln\mathcal{L}(\boldsymbol{\theta}, \boldsymbol{\psi})\right\rangle\right|_{\left(\boldsymbol{\theta}_{\text {fid }}, \boldsymbol{\psi}_{\text {fid }}\right)}.
\end{split}
\end{align}
The next step consists of evaluating the above expression at $\left({\boldsymbol{\theta}}_{\text{bias}}, \boldsymbol{\psi}_{\text{fixed}}\right)$ (i.e., we are introducing model $M_2$). Similarly to the previous derivation, we now choose $\boldsymbol{\theta}_{\text{bias}}$ such that it is the value of $\boldsymbol{\theta}$ that maximizes the log-likelihood at a certain $\boldsymbol{\psi}_{\text{fixed}}$, in the same way as $\boldsymbol{\theta}_{\text {fid }}$ maximizes the $\log$-likelihood at a certain $\boldsymbol{\psi}_{\text {fid }}$. This means that both $\left.\left\langle\partial_{\theta_i} \ln\mathcal{L}(\boldsymbol{\theta}, \boldsymbol{\psi})\right\rangle\right|_{\left(\boldsymbol{\theta}_{\text {fid}}, \boldsymbol{\psi}_{\text {fid }}\right)}$ and $\left.\left\langle\partial_{\theta_i} \ln\mathcal{L}(\boldsymbol{\theta}, \boldsymbol{\psi})\right\rangle\right|_{\left(\boldsymbol{\theta}_{\text{bias}}, \boldsymbol{\psi}_{\text{fixed}}\right)}$ vanish.
We notice that here $\boldsymbol{\psi}_{\text{fid}}$ and $\boldsymbol{\psi}_{\text{fixed}}$ have fixed values, thus in the above equations we have not maximized the full likelihood, but we have maximized the likelihood varying $\boldsymbol{\theta}$ only and considering $\boldsymbol{\psi}$ as a fixed set of parameters.
With this in mind, from Eq. (\ref{generalder}), we get

\begin{align}
\begin{split}
& \left.\Delta \theta_j\left\langle\partial_{\theta_i} \partial_{\theta_j} \ln\mathcal{L}(\boldsymbol{\theta}, \boldsymbol{\psi})\right\rangle\right|_{\left(\boldsymbol{\theta}_{\text {fid}}, \boldsymbol{\psi}_{\text {fid}}\right)}= \\
& -\left.\Delta \psi_j\left\langle\partial_{\theta_i} \partial_{\psi_j} \ln\mathcal{L}(\boldsymbol{\theta}, \boldsymbol{\psi})\right\rangle\right|_{\left(\boldsymbol{\theta}_{\text {fid }}, \boldsymbol{\psi}_{\text {fid}}\right)}
\end{split}
\end{align}
where the shifts are now evaluated at  $\Delta \theta_j=\theta_{\text{bias}, j}-\theta_{\text{fid}, j}$ and $\Delta \psi_j=\psi_{\text{fixed}, j}-\psi_{\text{fid}, j}$. By recalling that the averaged second derivatives of the likelihood are the Fisher matrices and taking into account Eq. (\ref{Vanish}), we notice that the above equation yields to

\begin{equation}\label{bias}
\Delta\theta_i=-\left(F^{\theta \theta}\right)_{i k}^{-1} F_{k j}^{\theta \psi} \Delta \psi_j.
\end{equation}

This result is the analogous of Eq. (\ref{1parameterbias}) in a general multi-dimensional parameter space. It relates the bias in a \emph{set} of cosmological parameters to several individually-parametrized systematics or to a single multi-parameter systematic model, in terms of the Fisher matrix components.

\subsection{Inverting the Fisher bias relation}

As in the previous section, let $m$ be the total number of cosmological parameters (i.e., $\theta_i$ with $i\in [ 0,m ]$) and $n$ the total number of systematic parameters (i.e., $\psi_j$ with $j\in [ 0,n]$). Then, $F^{\theta \psi}$ is a rectangular $(m\times n)$-matrix where each column is that of the Fisher matrix containing both $\theta$ and $\psi$ elements and $F^{\theta\theta}$ is the ($m\times m$) Fisher matrix of model $M_2$ but with $\psi$ kept fixed (not varied). 

The next step is to invert Eq. (\ref{bias}) in order to obtain an expression that relates the systematic bias $\Delta \psi$ to quantities we know from the observations (i.e., $\Delta \theta$). The result is the following:
\begin{equation}\label{constr}
F^{\theta \psi}\Delta\boldsymbol{\psi}=-F^{\theta\theta}\Delta\boldsymbol{\theta}.
\end{equation}

For operational purposes, we remark that in the above equation, the Fisher matrices are evaluated at $\left(\theta_{\text {fid}}, \psi_{\text {fid}}\right)$. In principle, this equation is valid in the following cases:
\begin{itemize}
    \item The case where $m=n$ (general case of the above), for which Eq. (\ref{constr}) corresponds to a linear system of $m$ equations for $n=m$ variables. If $m=n=1$, we are considering the bias in a single cosmological parameter and investigating a systematic with a single parameter as well. All the terms in the above equations are thus scalars in this case.
    \item The case where $m>n$, for which Eq. (\ref{constr}) corresponds to an over-determined system of $m$ equations for $n$ variables.
    Although in general there is no unique solution, a comparison with observational priors and estimates can allow to possibly discard some solutions; in practice, by examining a range of systematic parameters implied by different biases, one can assess their impact on the data vector and determine whether they plausibly account for an observed tension.
    
\end{itemize}
The equation cannot be applied in situations where $m<n$ as Eq. (\ref{constr}) then corresponds to and undetermined system and therefore has no unique solution.

We note that the above equation can still be applied  
if a change of variables is needed. For example, it is common to express the lensing constraints in the $S_8-\Omega_{\rm m}$ plane, where $S_8 = \sigma_8\sqrt{\Omega_{\rm m}/0.3}$; similarly, in CMB analyses, $A_{\rm s}$ and the optical depth $\tau$ are the directly measured quantities, with $\sigma_8$ derived as $\sigma_8 \propto A_{\rm s}^{1 / 2} e^{-\tau}$ and for dark-energy analyses, constraints can sometimes be expressed in terms of $w_{\rm p}=w_0+\left(1-a_{\rm p}\right) w_a$. Upon considering a different set of parameters $\zeta_m$ (so changing variables from $\theta_m$), the new Fisher matrix would be 
\begin{equation}\label{change}
F^{\prime}_{mn}=\sum_{ij}\dfrac{\partial\theta_i}{\partial \zeta_m}\dfrac{\partial\theta_j}{\partial \zeta_n}F_{ij}
\end{equation}
and the bias in the new parameters would be related to the old one as 
\begin{equation}
  \delta \zeta_n =\frac{\partial \zeta_n}{\partial \theta_i} \delta \theta_i.
\end{equation}
By inserting Eq. (\ref{bias}) in the above, we obtain the following expression
\begin{equation}
\delta \zeta_n =-\frac{\partial \zeta_n}{\partial \theta_i}\left(F^{\theta \theta}\right)_{i k}^{-1} F_{k j}^{\theta \psi} \delta \psi_j,
\end{equation}
for the bias in the new set of parameters as a function of the systematics.

In the next section we restrict our study to some example cases where $m>n$. In particular, we will choose $\delta\psi$ to be one-dimensional and $\delta\theta$ three-dimensional.

\section{Implementation}\label{implementation}
\begin{figure}
    \centering
    \includegraphics[width=\columnwidth]{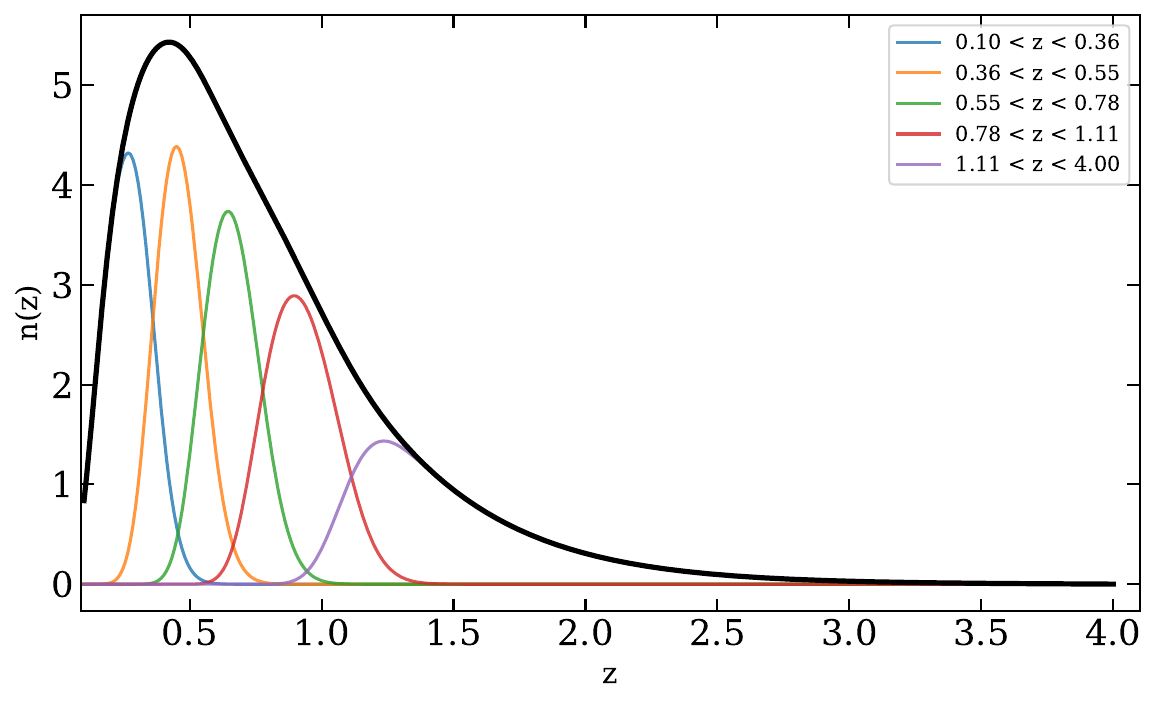}
    \caption{Redshift distribution $n(z)$ in the redshift range $z\in[0.1,4]$. The distribution has been obtained by splitting $ n(z)\propto z^2\exp\left(-(z/0.13)^{0.78}\right)\ $ into 5 equally-populated bins (see coloured lines) and convolving it with a Gaussian distribution of standard deviation $\sigma_z=0.05$ (black line).}
    \label{fig:redshift}
\end{figure}

To showcase the usefulness of our methodology, we create a cosmic shear synthetic data vector of a survey analogous to 1 year of observations from the NSF-DOE Vera C. Rubin Observatory's Legacy Survey of Space and Time (LSST). For the setup, we use emulate the choices made bythe Dark Energy Science Collaboration (DESC) in \cite{SRD} and assume a galaxy redshift distribution given by
\begin{equation}
    n(z)\propto z^2\exp\left(-(z/z_0)^\alpha\right)\,,
    \label{eq:SRD_redshift}
\end{equation}
with $(\alpha, z_0)=(0.13,0.78)$. This distribution is then split into 5 equally-populated bins and convolved with a Gaussian distribution of standard deviation $\sigma_z(1+z)=0.05(1+z)$. The final distribution can be seen in Fig. \ref{fig:redshift}. 
We generate a mock data vector using angular power spectra as our choice of observable, computing both auto- and cross-correlations between redshift bins. We use the Limber approximation \citep{Limber} and compute the angular power spectra as
\begin{equation}
    C^{(ij)}(\ell)=\int_0^{\chi_\mathrm{hor}} \mathrm{d}\chi \frac{W_\mathrm{G}^{(i)}(\chi)W_\mathrm{G}^{(j)}(\chi)}{\chi^2}P_\mathrm{m}\left(\frac{\ell+1/2}{\chi},z(\chi)\right)\,,
\end{equation}
where $P_\mathrm{m}$ is the matter power spectrum, $\chi_\mathrm{hor}$ the comoving radial distance to the horizon, and $W_\mathrm{G}^{(i)}(\chi)$ the lensing kernel for redshift bin $i$, given by
\begin{equation}
    W_\mathrm{G}^{(i)}(\chi)=\frac{3H_0^2\Omega_m}{2c^2}\frac{\chi}{a(\chi)}\int_\chi^{\chi_\mathrm{hor}} \mathrm{d}\chi' n^{(i)}(\chi')\frac{\chi'-\chi}{\chi'}\,,
\end{equation}
where $n^{(i)}$ is the galaxy redshift distribution, $a(\chi)$ the scale factor, $c$ the speed of light and $H_0$ the Hubble constant. 

We compute angular power spectra using the Core Cosmology Library ({\tt CCL}, \citealt{Chisari19}, version 3.0.2) -- the standard analysis package used by DESC. The matter power spectrum is obtained using CAMB (which evaluates the nonlinear matter power spectrum using halofit).
We also model the covariance of our synthetic data vector analytically, limiting the analysis only to Gaussian terms. Although current and future cosmic shear analyses require going beyond this assumption, it is sufficient in our case to illustrate the application of this methodology. The covariance is obtained by
\begin{equation}
\begin{split}
    C_A^{(ijkl)}(\ell)=&\frac{\delta_{\ell \ell'}}{2f_\mathrm{sky}\ell\Delta\ell}\,\times \\
    & \left(\bar{C}^{(ik)}(\ell)\bar{C}^{(jl)}(\ell)+\bar{C}^{(il)}(\ell)\bar{C}^{(jk)}(\ell)\right)\,,
\end{split}
\end{equation}
where $\Delta\ell$ is the width of the $\ell$ bin, $\delta_{ij}$ is the Kronecker symbol, $f_{\rm sky}$ the fraction of the sky coverage and we use the angular power spectra with the shape noise term added,
\begin{equation}
\bar{C}^{(ij)}(\ell)=C^{(ij)}(\ell)+\delta_{ij}\frac{\sigma^2_\epsilon}{2\bar{n}^i}\,.
\label{eq:noise_Cells}
\end{equation}
In the above, $\sigma_{\epsilon}$ is the 
intrinsic ellipticity dispersion and $\bar{n}^i$ is the mean angular number density of source galaxies in the $i$-th redshift bin (per steradian).
For the synthetic data we set $f_\mathrm{sky} = 0.4$ and $\sigma_{\epsilon}=0.26$ and assume a total effective number density of $n_\mathrm{eff}=10$ galaxies$/$arcmin$^2$ \citep{SRD}. We sample the data vector in 20 bins in the range $\ell \in [100, 2000]$ \citep[see for instance][]{Huang}. 

We can now use our synthetic data vector as our theoretical model and, together with its covariance, we can compute the Fisher information matrix from Eq. \eqref{Fishdef},
\begin{equation}
F_{ij}=\sum_\ell \left(\frac{\mathrm{d} \boldsymbol{D}^\mathrm{model}}{\mathrm{d} \theta_i}\right)^{\top}(\ell)\, C_A^{-1}(\ell)\frac{\mathrm{d} \boldsymbol{D}^\mathrm{model}}{\mathrm{d} \theta_j}(\ell)\,,
\end{equation}
where $\boldsymbol{D}^\mathrm{model}(\ell)$ is a vector containing all redshift bin combinations of $C^{(ij)}(\ell)$ and we omit writing out the indices of $C_A$. 
The Fisher matrix is computed by calculating numerical derivatives of the theoretical model with respect to the input parameters using a five-point stencil method. We have tested that the resulting Fisher matrix remains stable when we slightly vary the step size of our derivatives, ensuring robustness in the numerical derivative procedure.

As a demonstration, we designate three cosmological parameters to be the sampled parameters: $\theta=(\Omega_\mathrm{m}, A_\mathrm{s}, w_0)$, where $\Omega_{\rm m}$ is the matter density in the Universe today, $A_{\rm s}$ is the amplitude of the power spectrum, and $w_0$ is the equation of state parameter for dark energy (where $w_0=-1$ reduces to the $\Lambda$CDM model). In addition, we separately consider two systematic effects: intrinsic alignments and the effect of baryonic feedback on the matter power spectrum. The former affects our data vector linearly (for the model we consider) as it contributes as additive terms to the shear power spectrum, while the latter has a non-linear effect. Neglecting either of these systematics would result in biased constraints of the cosmological parameters, and we showcase how this can be diagnosed with our methodology. We expect that for intrinsic alignments our Fisher formalism will be more accurate but we also showcase the usefulness of the formalism for baryon feedback, a non-linear systematic.

\subsection{Intrinsic alignments}\label{Alignments}
Galaxies are subject to intrinsic alignments, which can cause significant biases in cosmological constraints from weak lensing measurements if left unaccounted for. Their effect is to modify the observed angular power spectra by adding three additional correlation terms,
\begin{equation}
    C^{(ij)}_\mathrm{obs}(\ell)=C^{(ij)}(\ell)+C^{(ij)}_\mathrm{GI}(\ell)+C^{(ij)}_\mathrm{IG}(\ell)+C^{(ij)}_\mathrm{II}(\ell)\,,
    \label{eq:IA_Cells}
\end{equation}
where the additional angular power spectra are given by
\begin{align}
    C^{(ij)}_\mathrm{GI}(\ell)&=\int_0^{\chi_\mathrm{hor}} \mathrm{d}\chi \frac{W_\mathrm{G}^{(i)}(\chi)n^{(j)}(\chi)}{\chi^2}P_\mathrm{\delta I}\left(\frac{\ell+1/2}{\chi},z(\chi)\right)\,, \\
    C^{(ij)}_\mathrm{II}(\ell)&=\int_0^{\chi_\mathrm{hor}} \mathrm{d}\chi \frac{n^{(i)}(\chi)n^{(j)}(\chi)}{\chi^2}P_\mathrm{II}\left(\frac{\ell+1/2}{\chi},z(\chi)\right)\,, 
\end{align}
and analogously for $C^{(ij)}_\mathrm{IG}(\ell)$. The power spectra in the above equations depend on the choice of the intrinsic alignment model. Here we use the non-linear linear alignment model \citep[NLA,][]{HirataS2004,BridleK2007}, resulting in
\begin{align}
    P_\mathrm{\delta I}\left(k,z\right)&=-A_\mathrm{IA}\frac{C_1\rho_\mathrm{crit}\Omega_m}{D(z)}\,P_\mathrm{m}(k,z)\,, \\
    P_\mathrm{II}\left(k,z\right)&=A_\mathrm{IA}^2\left(\frac{C_1\rho_\mathrm{crit}\Omega_m}{D(z)}\right)^2\,P_\mathrm{m}(k,z)\,.
\end{align}
Here, $C_1$ is a constant that we fix to $5\times 10^{-14}(h^2M_\odot/\mathrm{Mpc}^{-3})^{-2}$ \citep{BridleK2007}, $\rho_\mathrm{crit}$ the critical matter density, $D(z)$ the growth factor normalised to 1 at $z=0$, and $P_\mathrm{m}$ is the \emph{non-linear} matter power spectrum. The amplitude $A_\mathrm{IA}$ typically depends on the galaxy sample, and dictates the strength of the alignment signal. When constructing a synthetic data vector with intrinsic alignments, we use Eq. \eqref{eq:IA_Cells} both for the data vector, and for computing the covariance in Eq. \eqref{eq:noise_Cells}. 

\subsection{Baryonic feedback}\label{baryonfeed}
Weak lensing analyses can be heavily influenced by an incomplete understanding of the effect of supernovae feedback, gas cooling and active galactic nuclei (AGN) accretion processes on the matter distribution in the Universe. These phenomena are collectively referred to as \emph{baryonic feedback}. 
In particular, baryonic feedback, can impact the non-linear matter power spectrum by up to $\sim 30\%$ (see e.g., \citealt{Chisari19}).
Therefore, understanding and modelling this process is crucial for extracting accurate cosmological information from weak lensing measurements.

In this work we consider the \emph{baryonic halo model}, typically analysed through hydrodynamical simulations (for the original HMCODE see \cite{Mead2015}, for HMCODE2020 see \cite{Mead2021}. 
Specifically, HMCODE2020 models baryonic feedback by modifying 
the concentration-mass relation of the halo profiles as a function of their gas fraction, which depends on the AGN feedback strength, parametrised by $\log(T_{\rm AGN})$.
 
Separately from the previous case, we provide an example of how to apply our method when cosmic shear data are in tension with external probes due to incorrect modelling of baryonic feedback. We showcase how this tension can be explained by exploring the feedback models. 

\section{Demonstration of systematics exploration}\label{exploration}
In this section, we provide two examples to showcase our method. In both cases we start by generating a synthetic data vector containing the effects of one of the systematics under considerations. 
We then employ a model that does not correctly capture the systematics and we sample the likelihood using {\tt Nautilus} \citep{nautilus}. This yields biased best-fit cosmological parameters that are in tension with the synthetic data. 
Then, using the Fisher bias formalism, we infer what value of the systematic parameter could explain this tension and showcase that the correct value is derived. The goal is to illustrate how this procedure could be used in a real experiment to explore the systematics parameter space without needing to run expensive likelihood samplers. 

\subsection{Intrinsic alignment}
In this section, we provide an example of how to use Eq. (\ref{constr}) considering intrinsic alignments as our source of systematic errors (the systematic parameter under consideration for this case is $\psi=A_{\rm IA}$ ). We aim at assessing if a mock tension among different analysis of a set of cosmological parameters, can be due to the difference in the underlying models. 

Specifically, for the two models we define weak lensing tracers, respectively, with and without alignments. 
We start by generating a synthetic data vector for a model with intrinsic alignments (i.e., we set $\psi_{\rm fid}=0.05$).
We then use a model that does not account correctly for the systematics (i.e., $\psi_{\rm fixed}=0$). Given the high precision of our simulated experiment, we choose this fiducial value for $A_\mathrm{IA}$ to keep the cosmological parameter bias low (approximately $1\sigma$).
We then sample the likelihood using {\tt Nautilus} and obtain the likelihood's maximum a posteriori using the Nelder-Mead optimisation. This yields a bias in the cosmological parameters.

\begin{figure}
    \centering
    \includegraphics[width=\columnwidth]{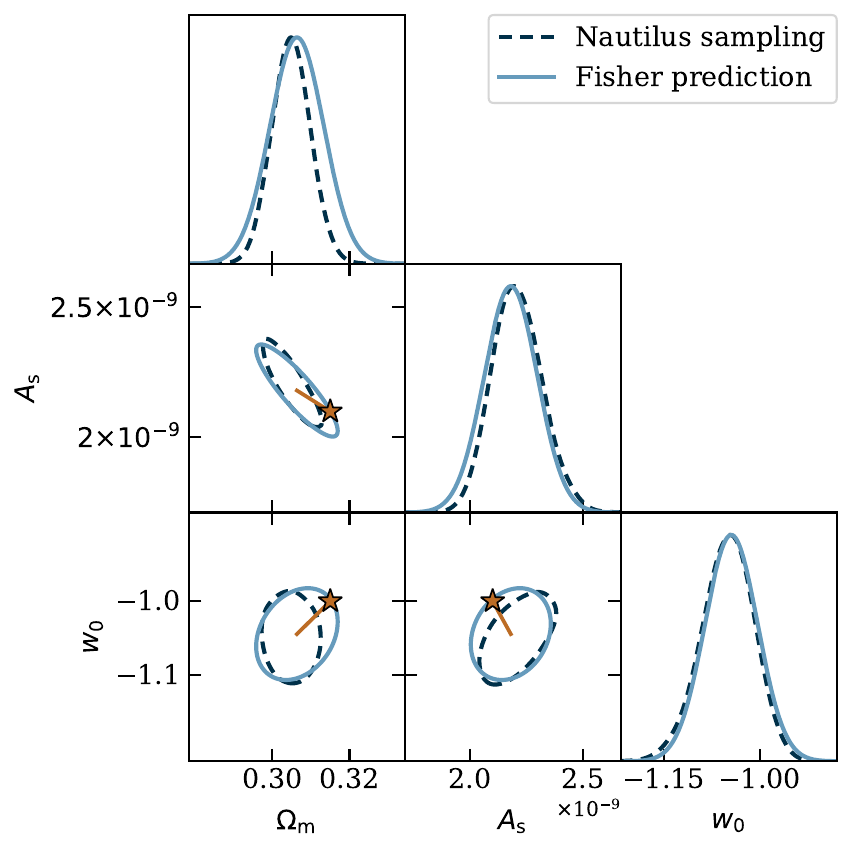}
    \caption{Fiducial constraints on the parameters of the $\Lambda$CDM model under consideration (i.e., $\Omega_{\rm m}$ and $A_{\rm s}$) and dark energy density ($w_0$) (marginalized over all the other parameters) from the Fisher formalism (solid lighter-blue line) compared against Nautilus nested sampler contours (dashed darker-blue line). The marginalized contours show the $68\%$ confidence level. The diagonal plots show 1D marginalized constraints.
    The shift from the best fit parameters (marked by stars) of the fiducial model (where the alignment amplitude is $A_{\rm IA}=0.05$) resulting from omitting IA is highlighted by the brown segment.}
    \label{fig:align_bias}
\end{figure}

Then, using our Fisher bias formalism, we infer what value of $A_{\rm IA}$ can explain the bias. 
To assess the validity of our analytic prediction, in Fig. \ref{fig:align_bias} we compare our result against the parameter shifts inferred from the full posterior sampled with \texttt{Nautilus}. The figure shows that the prediction from the Fisher formalism (solid lighter-blue line) is consistent with that obtained from the \texttt{Nautilus} nested sampler contours (dashed darker-blue line). 
This agreement is expected because the Fisher bias formalism relies on a first-order (linear) Taylor expansion around the fiducial model. Since the data vector depends linearly on $A_{\rm IA}$  and we chose a small offset $\Delta\psi$, the linear approximation remains valid, leading to consistent bias predictions in this regime.

We notice that Eq. (\ref{bias}) has been derived under the assumption that the chosen model does not depart significantly from the true one, and so that both $\delta \theta_i$ and $\delta\psi_j$ are \emph{small}. This means that our approach is accurate only as far as the confidence contours for the theoretical model and the fiducial one are \emph{close enough}. However, even in the cases were this condition is not satisfied, the formalism can provide a reliable indication of the magnitude and direction of the shift, possibly allowing to exclude certain systematics as the cause of a tension among different data-sets. 
 
The methodology we propose can be applied to a variety of alignment models, such as linear (LA) \citep[see][]{Catelan2001, HirataS2004, ChisariDv2013} and non-linear alignment (NLA) models \citep[see][]{HirataS2004, BridleK2007}, provided the likelihood is approximately Gaussian and can be linearized around the fiducial point (i.e., in the regime of small biases). Indeed, the non-linearity of the data vector does not affect the applicability of the method; it mainly controls how quickly the linear approximation breaks down as one moves away from the fiducial parameters. 
Although in some scenarios the IA contribution can mimic changes in cosmological parameters like $A_s$ and $\Omega_m$, this mostly affects how strongly the likelihood depends on the systematic parameters, without necessarily implying the breaking down of the approximation. In fact, more complicated non-linear models often include many poorly-constrained parameters, making the likelihood effectively more linear with respect to these parameters. 
Overall, the formalism works for any parametrized, nested systematics model, but care is required in interpreting the results and conclusions from this approach when the likelihood constrains the systematic parameters strongly (e.g. redshift uncertainties).

Our formalism could also be applied to perturbative models (TATT models) \citep[][]{BlazekVS2015, Blazek2019PhRvD, Schafer2009} that include additional, non-linear terms on top of the linear ones. 
These terms correspond to the effects of \emph{tidal torquing} and \emph{density-weighting}, and describe the torquing of a galaxy angular momentum by the surrounding tidal field and the additional weighting generated by the local density of galaxies, respectively.
As an example, the impact of neglecting the modelling of tidal torquing could be assessed by using the formalism developed here and fixing the corresponding systematic parameters in that model.
The formalism would allow to explore also NLA models including redshift evolution: given a certain $A_{\rm IA}$, one could consider a missing systematic parameter which comes from expanding the redshift evolution \citep[see e.g.,][]{Samuroff,DESY3Abbott2022} to first order.

\subsection{Baryonic Feedback}
\begin{figure}
    \centering
    \includegraphics[width=\columnwidth]{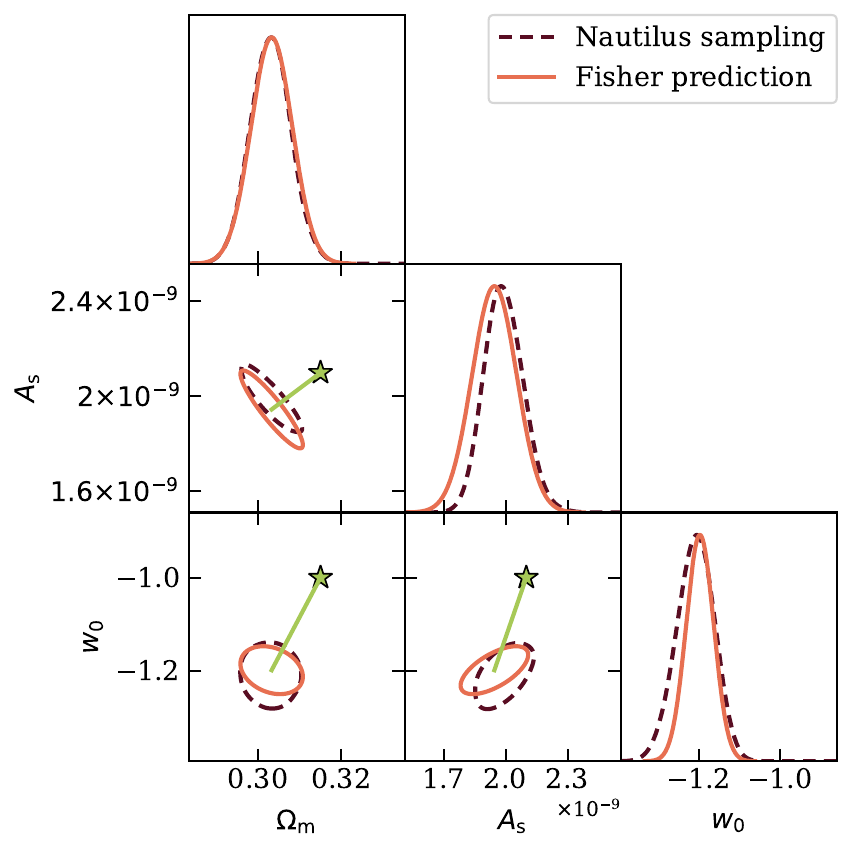}
    \caption{Fiducial constraints on $\Omega_{\rm m}$, $A_{\rm s}$ and $w_0$ (marginalized over all the other parameters) from the Fisher formalism (solid lighter line) compared against Nautilus contours (dashed darker line). The marginalized contours show the $68\%$ confidence level. The diagonal plots show 1D marginalized constraints. The green segment highlights the bias among the biased model, used to obtain our Fisher predictions, and the fiducial one (whose fiducial parameters are marked by the green stars). They differ in the value of $\log(T_{\rm AGN})$, such that $\delta (\log(T_{\rm AGN})) =8.0-7.7=0.3$.}
    \label{fig:triangle_agn}
\end{figure}

We now provide an example of how to use Eq. (\ref{bias}) in the case where both underlying models include baryonic feedback as systematics (i.e., $\psi=\log(T_{\rm AGN})$) but with different values. We follow a similar procedure to the previous example, but this time the fiducial model includes strong AGN feedback ($\log(T_{\rm AGN})=8$), while the (wrong) assumed model includes weaker AGN feedback ($\log(T_{\rm AGN})=7.7$). 
We again sample the likelihood using {\tt Nautilus}, identify its maximum a posteriori with the Nelder–Mead algorithm and evaluate the offset from the fiducial parameter values, thus extracting the bias $\Delta\theta$.
Then, through the Fisher bias relation, we estimate what value of $\log(T_{\rm AGN})$ could explain such bias. We compare our analytic predictions to the numerical sample in Figure \ref{fig:triangle_agn}. 

In the figure, we notice that mis-modelling baryonic feedback for our forecasting scenario produces a significantly biased best-fit (outside the $1\sigma$ contour). Because the simulations use a highly-constraining configuration based on LSST Year 1 data with only four cosmological parameters varied and extend up to 
quite non-linear scales ($\ell=2000$), the small-scale signal becomes increasingly non-linear and sensitive to baryonic effects. As a result, our chosen shift in $\log(\rm T_{AGN})$ leads to a significant bias in the best-fit. 

We also compare our analytic predictions with the parameter values inferred using {\tt Nautilus}. Fig. \ref{fig:triangle_agn} shows that the biased Fisher forecast aligns closely with the sampled posterior contours, confirming the robustness of the approach even in a case where the biased result is far from the fiducial value (compared to the accuracy of the experiment). 
We remark that the formalism is fully general and can be analogously applied to analysis performed with more refined models, e.g., based on the FLAMINGO simulations \citep[see][]{Schaye2023}, which include emulators for generating power spectra \citep[see][]{Schaller2025,Schaller2025arXiv}, as a mechanism for modelling baryonic effects. 

\section{Discussion}\label{discussion}

In this work, we demonstrated a methodology based on the Fisher bias formalism to assess whether a systematic effect modelled by some parameter(s) could account for a discrepancy in inferred sampled/cosmological parameters between different data-sets. 
This method indicates the magnitude of the systematic parameter value required to reproduce the observed sampled parameter shift and allows probing whether such a value is reasonable based on prior knowledge of the systematic effect. We provide two pedagogical example cases, using intrinsic alignments and the effect of baryonic feedback as separate systematic effects that can lead to tensions between a weak lensing experiment and external data-sets. 

The formalism proposed in this paper can be applied to any type of nested models, i.e., where an extended model that includes an additional systematic effect can be reduced to the fiducial model. Some further examples, in the case of weak gravitational lensing, are redshift-dependent and non-linear intrinsic alignment models, or the effect of lensing magnification. 
In the latter, magnification is included or neglected in the modelling of galaxy number count correlations, entering, when included, as an extra angular power spectrum term in correlations with galaxy number counts. Neglecting magnification corresponds to a restricted (nested) model in which this additional term is set to zero by fixing the magnification bias parameter to $s(z)=0.4$ \citep[see e.g.,][]{DESKrause2021}. 

The strength of our methodology relies in its reduced computational cost compared to what is typically done in cosmological surveys when data discrepancies arise, i.e., running many expensive likelihood sampling algorithms to test the effect of extending the parameter space of several systematic effects. This makes our method particularly relevant during the blinding phases of cosmological surveys.
In fact, based on the blinded observables, systematic assessments are needed to determine how including or excluding a particular systematic might alter the inferred parameter estimates. 
Our method allows for a fast, broad, and thus more effective exploration of the systematic parameter space. It also allows one to quantify the relative importance of each effect, thus guiding modelling priorities before unblinding.

As mentioned above, there are two limitations of this approach. First, it is valid if the number of sampled parameters is equal or greater than that of systematic unmodelled parameters. Second, it is mathematically accurate in the limit of smooth and well-behaved likelihoods (linear in the data vector and the cosmological parameters) and small biases. Its reliability decreases faster for large biases and non-linear likelihood functions. 
We also remark that for non-Gaussian likelihoods, Fisher forecasts at higher orders \citep[implemented for instance using DALI, see e.g.,][]{Sellentin2014,Sellentin2015,Sellentin2016, Ryan2023}, would be more reliable. 
However, even in cases where our biased parameters do not perfectly match the fiducial values (e.g. the mock bias is not recovered within e.g. $1\sigma$), our formalism still yields mostly correct predictions if compared to the full likelihood sample (see e.g. Figure (\ref{fig:triangle_agn})). In general, our methodology always provides at least an indication of the direction (and possibly of the magnitude) of the bias introduced by a given systematic. Indeed, a shift in a markedly different direction would imply that other systematics should be investigated. So, information on the direction of the bias alone can potentially allow to rule out implausible systematic explanations efficiently.

\acknowledgements
\section*{Acknowledgements}
This publication is part of the project ``A rising tide: Galaxy intrinsic alignments as a new probe of cosmology and galaxy evolution'' (with project number VI.Vidi.203.011) of the Talent programme Vidi which is (partly) financed by the Dutch Research Council (NWO). CG is funded by the MICINN project PID2022-141079NB-C32. IFAE is partially funded by the CERCA program of the Generalitat de Catalunya.

\bibliography{bibliography}

\end{document}